\documentclass[11pt,twoside,a4paper,cmspaper,final,collab]{cms-tdr}
\def\svnVersion{fa094c2-D}\def\svnDate{2026/03/20}\def\cmsCernNoTag{CERN-EP-2025-186}\def\cmsCernDate{\today}\def\cmsMessage{Published in Physical Review D as \href{http://dx.doi.org/10.1103/w3b5-92ls}{\doi{10.1103/w3b5-92ls}.}}
\begin{document}\cmsNoteHeader{SUS-24-003}

\newlength{\cmsFigWidth}
\ifthenelse{\boolean{cms@external}}{%
    \providecommand{\cmsLeft}{upper\xspace}
    \providecommand{\cmsRight}{lower\xspace}
    \setlength{\cmsFigWidth}{\columnwidth}
}{%
    \providecommand{\cmsLeft}{left\xspace}
    \providecommand{\cmsRight}{right\xspace}
    \setlength{\cmsFigWidth}{0.6\textwidth}
}

\newcommand{\mll}{\ensuremath{m_{\ell\ell}}\xspace}
\newcommand{\mlt}{\ensuremath{m_{t\ell}}\xspace}
\newcommand{\mtautau}{\ensuremath{m_{\PGt\PGt}}\xspace}
\newcommand{\deta}{\ensuremath{\Delta\eta}\xspace}
\newcommand{\dphi}{\ensuremath{\Delta\phi}\xspace}
\newcommand{\tautau}{\ensuremath{\PGt\PGt}\xspace}

\newcommand{\PZstar}{{\HepParticle{\PZ}{}{\ast}}\xspace}
\newcommand{\PZGammaStar}{{\PZ/\HepParticle{\PGg}{}{\ast}}\xspace}

\newcommand{\wjets}{\ensuremath{\PW\text{+jets}\to\Pell\PGn\text{+jets}}\xspace}
\newcommand{\zjets}{\ensuremath{\PZ\text{+jets}\to\PGn\PGn\text{+jets}}\xspace}
\newcommand{\ztautau}{\ensuremath{\texorpdfstring{\PZGammaStar\to\TT}{Z->tau tau}}\xspace}

\newcommand{\PSGczDotTf}{{\HepParticle{\PSGc}{1,2,3,4}{0}}\xspace}
\newcommand{\PSGcpmDot}{{\HepParticle{\PSGc}{1,2}{\pm}}\xspace}
\newcommand{\PGrz}{{\HepParticle{\uprho}{}{0}}\xspace}
\newcommand{\PQjo}{{\HepParticle{j}{1}{}}\xspace}

\newcommand{\Phase}[1]{\mbox{Phase #1}\xspace}

\newcommand{\Dmz}{\ensuremath{\Delta m^0}\xspace}
\newcommand{\Dmpm}{\ensuremath{\Delta m^\pm}\xspace}
\newcommand{\BDT}{\ensuremath{\text{BDT}}\xspace}

\cmsNoteHeader{SUS-24-003}
\title{Search for Higgsinos in final states with low-momentum lepton-track pairs at \texorpdfstring{13\TeV}{13 TeV}}

\date{\today}

\abstract{We present a search for the pair production of Higgsinos in final states with large missing transverse momentum and either two reconstructed muons or a reconstructed lepton (muon or electron) and an isolated track. The analyzed data correspond to proton-proton collisions with an integrated luminosity of 137\fbinv, collected by the CMS experiment at $\sqrt{s}=13\TeV$ in 2016, 2017, and 2018. The signal scenario assumes four nearly mass degenerate higgsino mass eigenstates: two neutralino states \PSGczDt and \PSGczDo with a small mass difference in the range 1--10\GeV and two chargino states \PSGcpmDo with an intermediate mass. The analysis focuses on the decay of the heavier neutralino into the lighter one and a virtual \PZ boson, which decays into two same-flavor leptons. The leptons have small transverse momentum and/or a small opening angle between the identified muons. An isolated track is used to recover events in which only one of the two leptons is identified. Multivariate discriminants are used to enhance the sensitivity by efficiently rejecting backgrounds from SM processes or misreconstructed tracks and/or leptons. The search explores a unique phase space and probes a previously unexplored region of the signal model parameter space. Mass differences between the two neutralinos are probed down to 1.5\GeV, assuming a Higgsino mass of 100\GeV. The maximum excluded Higgsino mass is 115\GeV.}

\hypersetup{%
pdfauthor={CMS Collaboration},%
pdftitle={Search for Higgsinos in final states with low-momentum lepton-track pairs at 13 TeV},%
pdfsubject={CMS},%
pdfkeywords={CMS, supersymmetry}}

\maketitle

\section{Introduction}
\label{sec:search-introduction}

Physics scenarios beyond the standard model (SM) that feature near-degenerate electroweak (EW) doublets or multiplets are well-motivated candidates for explaining dark matter (DM)~\cite{Zwicky:1937zza,Rubin:1970zza}.
For example, in inert doublet~\cite{LopezHonorez:2006gr} and inelastic DM~\cite{Tucker-Smith:2001myb,Izaguirre:2015zva} models, the lightest of two or more states, if neutral and stable, may account for the observed DM relic abundance.
Higgsinos in $R$-parity-conserving supersymmetry (SUSY)~\cite{Ramond:1971gb,Golfand:1971iw,Neveu:1971rx,Volkov:1972jx,Wess:1973kz,Wess:1974tw,Fayet:1974pd,Fayet:1976cr,Nilles:1983ge} are also viable DM candidates, offering a solution not only to the DM puzzle but also to the large~\cite{Martin:1997ns} and small~\cite{BARBIERI198863,de_Carlos_1993} hierarchy problems, whereby fine-tuning among the Lagrangian parameters remains unexplained.
To address these issues, Higgsino mass eigenstates must typically be of the order of 100\GeV, making them potentially detectable at the CERN LHC.
The viable parameter space of these models has been explored and constrained by searches for beyond-the-SM physics using the \mbox{Run 2} (2016--2018) data of ATLAS~\cite{ATLAS:2024qxh} and CMS~\cite{CMS:2024gyw}, by experiments at the CERN Large Electron-Positron Collider (LEP)~\cite{Ask:2003pg}, and by DM direct detection experiments~\cite{LZ:2022lsv,XENON100:2016sjq,PandaX:2014ria}.
Some parameter space remains unprobed, particularly in scenarios with highly compressed mass spectra, where ``highly compressed'' refers to mass splittings between produced and decayed particles of or less than the scale of the \PJGy.

In the minimal supersymmetric standard model (MSSM), the SM Lagrangian is extended to be invariant under SUSY transformations and includes a second complex scalar doublet.
A total of eight Higgs and Higgsino fields before EW symmetry breaking give rise to five physical Higgs bosons and four Higgsino mass eigenstates after symmetry breaking.
The Higgsinos mix with the superpartners of the \PW and \PB bosons (wino and bino) to form four chargino (\PSGcpmDot) and four neutralino (\PSGczDotTf) states, collectively referred to as electroweakinos.
In the limit where the wino and bino mass parameters are much larger than the Higgsino mass parameter, the lightest supersymmetric particle (LSP) emerges as the lightest state \PSGczDo among four nearly degenerate electroweakino states, which is stable.
Although small relative to the mass scale, a minimal mass difference of approximately 300\MeV between the chargino and the LSP is required by radiative corrections and cosmological constraints~\cite{Nagata:2014wma}.

\begin{figure*}[ht!]
  \centering
  \includegraphics[width=0.45\textwidth]{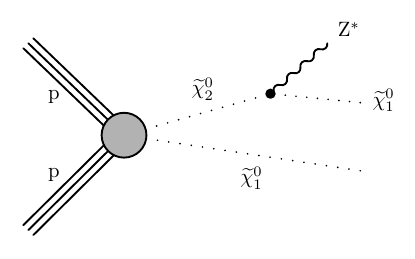}\hspace{0.05\textwidth}%
  \includegraphics[width=0.45\textwidth]{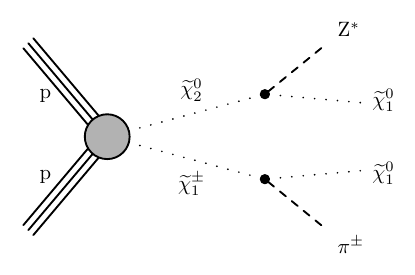}
\caption{Diagrams illustrating the production and decay of electroweakinos in the Higgsino simplified model via the $\PSGczDt\PSGczDo$ (\cmsLeft) and $\PSGczDt\PSGcpmDo$ (\cmsRight) processes.}
\label{fig:signal-feynman-diagrams}
\end{figure*}

We present a search for compressed-spectrum Higgsinos using proton-proton ($\Pp\Pp$) collision data recorded by CMS at a center-of-mass energy of 13\TeV, targeting events featuring the decay $\PSGczDt\to\Pell\Pell\PSGczDo$, where \Pell is a muon or electron.
The analysis focuses on the so-called natural subspace, characterized by large values of the ratio between the two vacuum expectation values, $\tan\beta$, and featuring a spectrum where the mass difference $\Dmz=m(\PSGczDt)-m(\PSGczDo)$ is twice the value of that between the chargino and the LSP, $\Dmpm=m(\PSGcpmDo)-m(\PSGczDo)=\Dmz/2$.
Two processes dominate the total signal production cross section and are illustrated in Fig.~\ref{fig:signal-feynman-diagrams}.
In both cases, the decay of the \PSGczDt usually proceeds via a virtual \PZ boson, which subsequently decays to a pair of low-momentum (soft) muons or electrons.
A branching fraction $\mathcal{B}(\PSGczDt\to\PGm\PGm\PSGczDo)=\mathcal{B}(\PSGczDt\to\Pe\Pe\PSGczDo)$ of 5\% is assumed, an approximate upper bound in the MSSM.
In the production mode featuring a chargino \PSGcpmDo, the \PSGcpmDo decays predominantly to hadrons, most often a single soft pion, via an off-shell \PW boson~\cite{Nagata:2014wma}, to which the analysis is insensitive.
The analysis object selection is not sensitive to these decays, and a branching fraction of 100\% for this dominant decay mode is assumed.
The search targets events with two opposite-charge, same-flavor leptons where the magnitude of the missing transverse momentum (\ptmiss) is large due to the two undetected LSPs with transverse momentum that recoils off of initial-state radiation (ISR).

Previous searches performed at the CERN LEP~\cite{ALEPH:2002gap,2004247,DELPHI:2003uqw,Acciarri_2000,OPAL:2003nhx,OPAL:2002lje} exclude charginos with masses up to approximately 90--100\GeV.
The ATLAS and CMS Collaborations have extended the exclusion reach to 205\GeV for a mass splitting \Dmpm of 7.5\GeV, using final states with two reconstructed, isolated leptons~\cite{Aad_2020,ATLAS:2021moa,CMS:2021edw}.
However, the sensitivity decreases for smaller values of \Dmpm and becomes negligible for splittings below 3\GeV, primarily due to kinematic and isolation-based selection requirements on the leptons, such as $\pt>3.5\GeV$ and angular separation $\DR(\Pell_1, \Pell_2)>0.3$~\cite{CMS:2021edw}.
For \Dmpm values below 0.7\GeV, sensitivity is recovered by searches targeting charginos with macroscopic lifetimes using soft and isolated, as well as disappearing track signatures~\cite{CMS:2020atg,ATLAS:2022rme,CMS:2023mny,ATLAS:2024umc}.
Nonetheless, a gap in sensitivity remains between the regimes targeted by track-based and soft lepton searches.

To improve sensitivity in this region, we analyze events with lepton candidates selected using relaxed kinematic thresholds and identification criteria compared to previous analyses.
Signal regions (SRs) are constructed using boosted decision trees (BDTs) trained to distinguish signal-like events in the target phase space.
The SRs are mutually exclusive and statistically independent, both with respect to each other and to those used in previous CMS analyses.
Three event categories are considered:
\begin{itemize}
\item dimuon: events with two reconstructed and identified muons;
\item muon + exclusive track: one reconstructed muon and an isolated exclusive track;
\item electron + exclusive track: one reconstructed electron and an isolated exclusive track.
\end{itemize}
Here, ``exclusive'' means that the track is not geometrically matched to any identified lepton and is used as a proxy to recover leptons that were not identified.
This category recovers up to 50\% of such lost leptons, but contributes only marginally to the final results because of residual background.
The use of events with two identified muons where either at least one of the muons has $\pt<3\GeV$ or $\DR(\Pell\Pell)<0.3$, as well as the lepton+exclusive track categories, make this analysis the first of its kind, establishing sensitivity to previously unexplored regions of parameter space. A study with a similar final state has been presented by the ATLAS Experimen~\cite{ATLAS:2025lhc}.

The paper is structured as follows.
Section~\ref{sec:detector} provides a brief description of the CMS detector, event reconstruction, and simulation.
The object and event selection procedures are described in Section~\ref{sec:object-selection}.
The background estimation methodology and the assessment of systematic uncertainties are presented in Section~\ref{sec:background-estimation}.
Section~\ref{sec:results} contains the results of the search, and a final summary can be found in Section~\ref{sec:summary}.
Tabulated results are provided in the HEPData record for this analysis~\cite{hepdata}.

\section{The CMS detector, event reconstruction, and simulation}
\label{sec:detector}

The CMS apparatus~\cite{CMS:2008xjf,CMS:2023gfb} is a multipurpose, nearly hermetic detector, designed to trigger on~\cite{CMS:2020cmk,CMS:2016ngn,CMS:2024aqx} and identify electrons, muons, photons, and charged and neutral hadrons~\cite{CMS:2020uim,CMS:2018rym,CMS:2014pgm}.
A global ``particle-flow'' (PF) algorithm~\cite{CMS:2017yfk} aims to reconstruct all individual particles in an event by combining information from the all-silicon inner tracker, crystal electromagnetic and brass-scintillator hadron calorimeters, which operate inside a 3.8\unit{T} superconducting solenoid, and the gas-ionization muon detectors embedded in the flux-return yoke outside the solenoid.
The reconstructed particles are used to build leptons and jets, as well as \ptmiss~\cite{CMS:2016lmd,CMS:2019ctu}.

The tracking system, which plays a central role in this analysis, consists of silicon pixel and strip detectors located within the solenoid volume~\cite{CMS:2008xjf,CMSTrackerGroup:2020edz}.
The inner tracker used during the 2016 data-taking period, referred to as the ``\Phase0'' tracker, measured charged particles within $\abs{\eta}<2.5$.
An upgraded pixel detector, known as the ``\Phase1'' tracker, was installed at the beginning of 2017 and used for the 2017 and 2018 data-taking periods.
The \Phase1 tracker extended the coverage to $\abs{\eta}<3.0$.
In the barrel region, charged-particle tracks pass through three (four) pixel layers within a radius of 102 (160)\mm in the \Phase0 (\Phase1) tracker.
The strip tracker provides up to ten additional tracking layers within a radius of 1.2\unit{m}.
Compared to the \Phase0 tracker, the \Phase1 upgrade improves both tracking and vertex reconstruction performance and enhances the efficiency of algorithms identifying displaced \PQb jets.

The electromagnetic calorimeter (ECAL) is a total absorption calorimeter that measures the energy of electromagnetic showers using lead tungstate crystals.
The lead tungstate crystals are designed to fully contain energetic electromagnetic showers and provide high resolution.
Because of the 3.8\unit{T} magnetic field from the CMS solenoid, electromagnetic showers from electrons and photons are broadened in the azimuthal angle and narrowed in the polar angle.
These characteristics are used to distinguish showers from electrons and photons from those of hadrons.

The muon system comprises gas-ionization detectors embedded in the steel flux-return yoke outside the solenoid, enabling muon reconstruction within $\abs{\eta}<2.4$.
It consists of three types of detectors: drift tube chambers in the central region, cathode strip chambers in the forward region, and resistive-plate chambers covering both regions.
The fine granularity of the endcap muon detectors provides efficient reconstruction for low-\pt muons in the high-$\abs{\eta}$ region~\cite{CMS:2018rym}, which are important for this analysis.
In contrast, muon acceptance is lower in the central region due to the significant curvature of low-\pt charged particles in the magnetic field~\cite{CMS:2018rym}.

Electron tracks are reconstructed using a Gaussian sum filter~\cite{Adam:2003kg} that accounts for measurement errors, energy loss, and multiple scattering as the electron traverses the detector volume.
The Gaussian-sum-filter tracks are matched with energy clusters in the ECAL to form electron objects~\cite{CMS:2015xaf}.
Electrons that pass a loose set of selection criteria are used in this analysis because of their high reconstruction efficiency.

Jets are reconstructed offline from PF candidates clustered using the anti-\kt algorithm~\cite{Cacciari:2008gp,Cacciari:2011ma} with a distance parameter of 0.4.
Raw jet energies are corrected to establish a uniform calorimeter response as a function of $\eta$ and a calibrated absolute response in transverse momentum \pt.
Jets arising from the hadronization of \PQb quarks, referred to as \PQb jets, are identified by the combined secondary vertex algorithm based on deep neural networks, \textsc{DeepCSV}~\cite{Sirunyan:2017ezt}.
This algorithm achieves a high identification efficiency of around 80\%.

{\tolerance=800
The SM production of \wjets, \zjets, QCD, and \ttbar events is simulated using the {\MGvATNLO}\,2.2.2~\cite{Alwall:2014hca,Alwall:2007fs} event generator at leading order (LO) precision.
The \ttbar events are generated with up to three additional partons in the matrix element calculations.
The \wjets and \zjets events are generated with up to four additional partons.
Diboson events, such as those originating from $\PW\PW$ and $\PZ\PZ$ production, are generated with {\MGvATNLO}\,2.2.2 at next-to-leading order (NLO)~\cite{Frederix:2012ps}, except that $\PW\PW$ events in which both \PW bosons decay leptonically are generated using the {\POWHEG}\,v2.0~\cite{Nason:2004rx,Frixione:2007vw,Alioli:2010xd,Alioli:2009je,Re:2010bp} program at NLO.
For background events, \Phase0 samples use the CUETP8M1 tune~\cite{Khachatryan:2015pea}, while \Phase1 samples use the CP5 tune~\cite{CMS:2019csb}.
Parton showering and hadronization are simulated using the {\PYTHIA}\,8.205 generator~\cite{Sjostrand:2014zea}.
The detector response is modeled with the \GEANTfour~\cite{Agostinelli:2002hh} suite of programs. Additional proton-proton collisions, distinct from the primary vertex (PV) but occurring in the same bunch crossing, are emulated at LO and superimposed onto the event.
\par}

Signal samples are generated using the CP2 tune~\cite{CMS:2019csb}.
Samples generated at LO (NLO) with the CUETP8M1 tune use the NNPDF3.0LO (NNPDF3.0NLO) parton distribution functions (PDFs)~\cite{Ball:2013hta}, while those generated with the CP2 or CP5 tune use the NNPDF3.1LO (NNPDF3.1 NNLO) PDFs~\cite{Ball:2017nwa}.
Simulated signal events are generated at leading order (LO) using the {\PYTHIA}\,8.205 generator~\cite{Sjostrand:2014zea}.
To manage computational resources, the detector response for signal events is modeled using the CMS fast simulation framework~\cite{Abdullin:2011zz,Giammanco:2014bza}, which yields results generally consistent with those obtained from \GEANTfour.
To improve agreement with \GEANTfour, a correction of 1\% is applied to account for differences in the efficiency of the jet quality requirements~\cite{CMS:2016lmd,CMS:2019ctu}, and corrections of 5--12\% are applied to account for differences in the \PQb tagging efficiency.

\section{Object and event selection}
\label{sec:object-selection}

Signal event candidates for this analysis are recorded using triggers that require \ptmiss to exceed a threshold between 100 and 120\GeV, depending on the instantaneous luminosity of the LHC.

Muons are selected with \pt between 2 and 15\GeV and $\abs{\eta}<2.4$.
Electrons are selected with \pt between 5 and 15\GeV and $\abs{\eta}<2.5$.
Quality criteria are applied to the track fit and to the consistency of momentum measurements across subdetectors, aiming to optimize the balance between selection efficiency and the misidentification rate. No requirements are placed on the impact parameter of reconstructed leptons with respect to the PV, to retain full efficiency in cases where leptons are displaced by up to several millimeters from the PV.
Both muons and electrons are required to have $\DR>0.4$ from the leading jet and to satisfy the jet-based isolation requirement described below.

Jets are selected with $\pt>15\GeV$ and $\abs{\eta}<5.0$.
Two definitions of missing transverse momentum are used: the standard \ptmiss and an alternative referred to as hard \ptmiss.
For the hard \ptmiss the transverse momentum vector sum is performed over all objects with $\pt>30\GeV$ instead of over all reconstructed particles in the event.
Unlike the standard \ptmiss, the hard \ptmiss is uncorrelated with the jet-based lepton isolation variable used to define the signal and control regions (CRs), as explained below, and is more robust against potential mismodeling of simultaneous proton-proton collisions in a given bunch crossing (pileup).

A custom jet-based isolation criterion is defined for leptons using a corrected set of jets.
Jets are selected from all reconstructed jets with $\pt<30\GeV$.
For each such jet, the momentum of each selected lepton within the jet is vectorially subtracted from the jet momentum, resulting in the corrected jet.
The isolation criterion is determined by two parameters: the lower threshold on the \pt of the nearest corrected jet and the upper threshold on the \DR between the lepton and that jet.
If the lepton does not lie within a \DR of 0.6 of a corrected jet with $\pt>10\GeV$, it is considered isolated.

Tracks are required to have $\pt>1.9\GeV$ and $\abs{\eta}<2.4$.
The relative isolation is defined as the ratio of the scalar sum of \pt of other tracks within a cone of radius 0.3 around the candidate track to the track \pt, and must be less than 0.1.
Tracks must have transverse and longitudinal impact parameters $\abs{d_{xy}}<0.02\cm$ and $\abs{d_{z}}<0.02\cm$, with respect to the PV, which is the reconstructed vertex with the largest value of summed physics-object $\pt^2$.
No selected muon or electron may lie within \DR of 0.01 of the track.
Tracks satisfying these criteria are referred to as exclusive tracks.

Dedicated track-picking BDT classifiers are used to identify which track in each signal event corresponds to the lepton from the neutralino decay.
Four separate BDTs are trained, corresponding to the two lepton flavors and the two detector phases.
All classifiers share a common structure of 200 decision trees with a maximum depth of 3.
They are trained using the \textsc{AdaBoost} algorithm and the Gini index as the separation criterion, as implemented in the \textsc{Hocker:2007ht} package~\cite{Therhaag:2010zz}.

Training is performed on tracks from a dedicated signal simulation, selected using the same object criteria as the analysis but with slightly loosened requirements: a track $\pt>1\GeV$ and no vetoes based on the invariant mass of the dilepton system, which are otherwise applied in the preselection.
The preselection described below is applied to the simulated events used in the training samples, which are split randomly and evenly between training and test sets.
A broad range of Higgsino mass parameters corresponding to $m(\PSGcpmDo)$ is considered, spanning 100--500\GeV.
The training is restricted to models with mass splittings \Dmz in the range 0.3--4.6\GeV, prepared in \Dmz steps of 100\MeV.
Tracks are labeled as signal if they originate from leptons in the $\PSGczDt\to\PSGczDo\Pell^+\Pell^-$ decay, as determined by geometrical matching of trajectories, and as background otherwise.
The training and test sets are compared to ensure there is no overtraining.

The jet-based isolation variable is uncorrelated with the hard \ptmiss because the soft jets used to define it fall below the \pt threshold used in the computation of hard \ptmiss.
Leptons that fail the isolation requirement define a sideband CR enriched in jet-related background.
Background contributions from this region are estimated as described in Section~\ref{sec:background-estimation}.

\subsection{Event selection}

\begin{figure}[!ht]
\centering
  \includegraphics[width=0.45\textwidth]{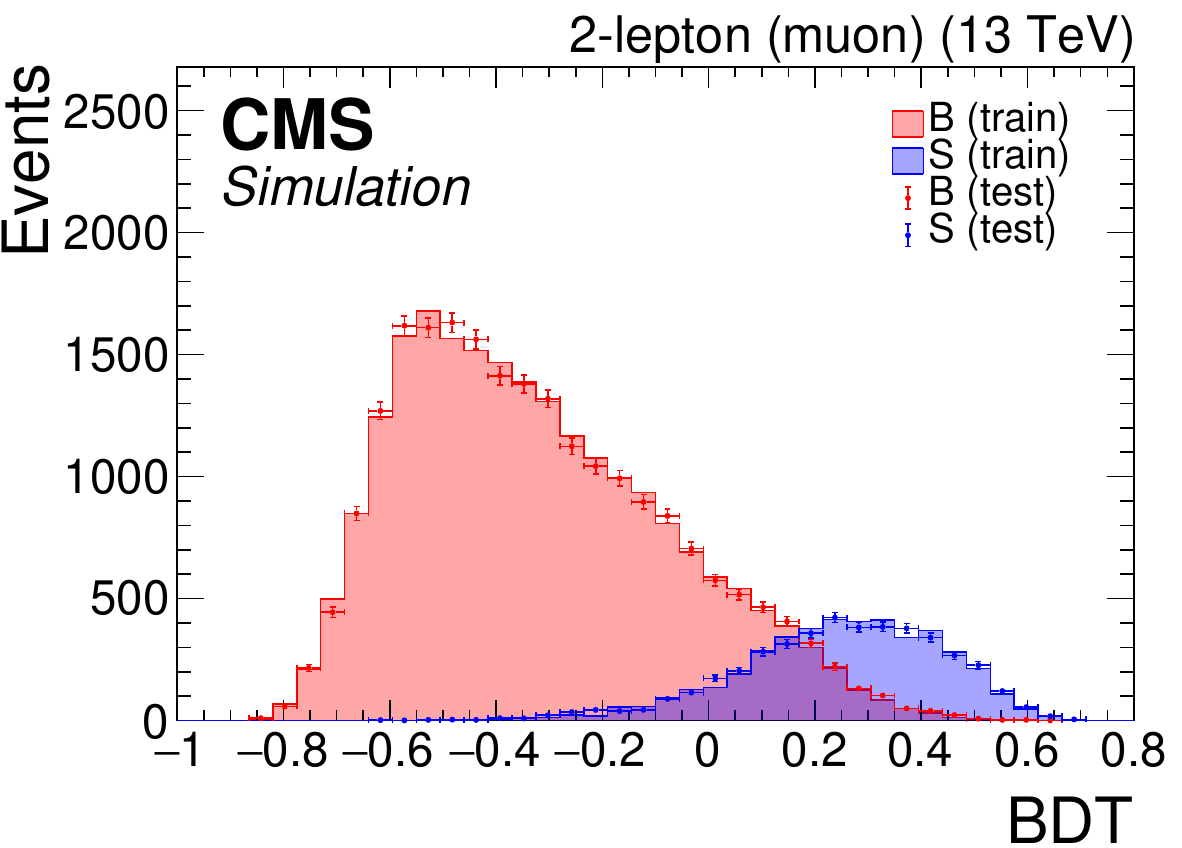}
  \includegraphics[width=0.45\textwidth]{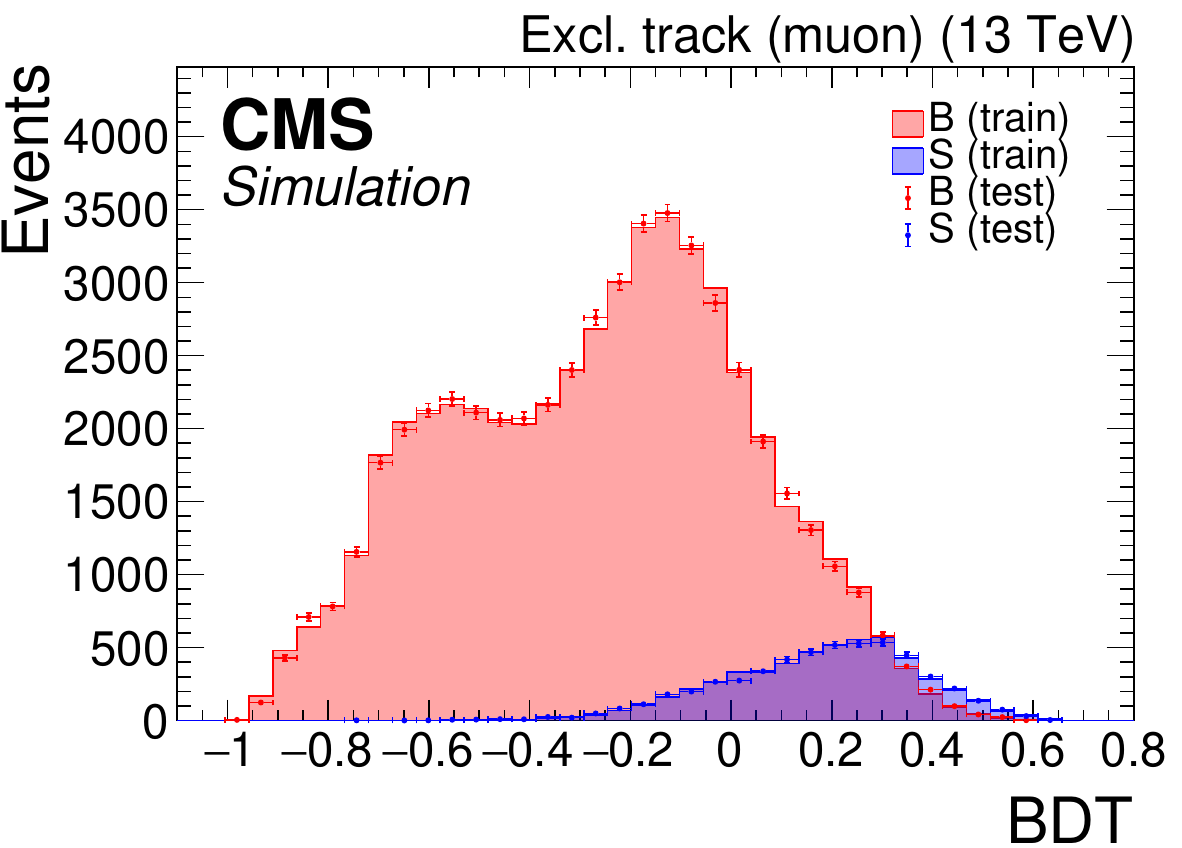}
\caption{Unweighted distributions of the event-level BDT scores for events from the signal and background training samples in the dimuon category (\cmsLeft) and the muon+track category (\cmsRight), based on the \Phase1 detector configuration.}
\label{fig:event-bdt-dimuon-output}
\end{figure}

The event preselection, common to all analysis categories, consists of the following requirements:
\begin{itemize}
\item hard $\ptmiss>220\GeV$ and $\ptmiss>140\GeV$, to select events efficiently with respect to both signal acceptance and trigger thresholds;
\item at least one jet with $\pt>30\GeV$ and $\abs{\eta}<2.4$, consistent with the presence of ISR;
\item $N_{\PQb\text{ jets}}=0$, namely zero reconstructed \PQb jets, to suppress \ttbar background;
\item $\min{\dphi\big(\text{hard }\ptvecmiss,\ptvec\big)}>0.4$ for all jets to reduce events with mismeasured jets contributing to hard \ptmiss;
\item $N_{\Pell}^{\text{hard}}=0$, to veto isolated leptons with $\pt>30\GeV$, suppressing backgrounds from $\PW\to\Pell\PGn$ decays;
\item $0.4<\mll<12\GeV$, to target the compressed mass region of interest.
\end{itemize}
In the dimuon category, two reconstructed and identified muons are required, and events must satisfy the following criteria:
\begin{itemize}
\item $N_{\PGm}=2$, oppositely charged;
\item $\pt(\PGm_2)<3.5\GeV$ or $\DR(\PGm_1,\PGm_2)<0.3$, to ensure no overlap with the search described in Ref.~\cite{CMS:2021edw};
\item $\DR(\PGm_{1,2},\PQjo)>0.4$, where \PQjo is the leading jet;
\item \mll outside the ranges 0.75--0.81 and 3.0--3.2\GeV, to veto \PGo, \PGrz, and \PJGy resonances;
\item event $\BDT>0$, to enhance signal purity and reject SM backgrounds.
\end{itemize}
The dimuon event BDT is constructed using several observables, including the leading and subleading muon \pt, the \DR and $\deta$ between the two muons, the hard \ptmiss, and the differences in azimuthal angle between the hard \ptmiss and the muons, as input features. The full set of features is listed in Table~\ref{tab:dimuon-bdt-variables}, ordered by their relative importance.
The most discriminating variable is the invariant mass of the dimuon system, which has the characteristic endpoint for signal corresponding to $\Dmz$ compared to background. The dimuon event BDT output distributions are shown in Fig.~\ref{fig:event-bdt-dimuon-output}.
Six SRs are defined for events with BDT output scores in bins with edges of 0, 0.1, 0.2, 0.3, 0.4, 0.5, and 1, given in order of increasing sensitivity.
A single event BDT is used for both detector configurations. The lepton + exclusive track event-level BDT importance rankings are similar, with the \DR being most highly ranked.

\begin{table}[!ht]
\centering
\topcaption{Input variables to the event-level BDT used in the dimuon category, ranked by their importance in descending order. The symbols $\ell_1$ and $\ell_2$ denote the leading and subleading leptons, while the variable $m_{T}$ refers to the transverse mass~\cite{Arnison:1983rp}. The minimization in $\min\dphi$ is over jets.}
\label{tab:dimuon-bdt-variables}
\begin{scotch}{ccl}
&Rank & Variable \\ \hline
&1 & \mll \\
&2 & $\pt(\ell_1)$ \\
&3 & $\text{hard }\ptmiss$ \\
&4 & $H_\mathrm{T}$ \\
&5 & $\DR(\ell_1,\ell_2)$ \\
&6 & $\min\dphi(\text{hard }\ptvecmiss,\ptvec(j))$ \\
&7 & $\pt(\ell_1+\ell_2)$ \\
&8 & $\pt(\PQjo)$ \\
&9 & $\pt(\ell_2)$ \\
&10 & $\eta(\ell_1)$ \\
&11 & $m_\mathrm{T}(\ell_1)$ \\
&12 & $\abs{\dphi(\ell_2,\text{hard }\ptvecmiss)}$ \\
&13 & $\abs{\dphi(\ell_1,\text{hard }\ptvecmiss)}$ \\
&14 & $\abs{\dphi(\ell_1,\ell_2)}$ \\
&15 & $N_{\text{jets}}$ \\
&16 & $\eta(\PQjo)$ \\
&17 & $\abs{\deta(\Pell_1,\Pell_2)}$ \\
&18 & \mtautau \\
\end{scotch}
\end{table}

In the lepton + exclusive track categories, the track with the highest track-picking BDT score in each event is selected as the recovered lepton candidate.
Events in this category must satisfy the preselection and baseline selections, as well as the following additional criteria:
\begin{itemize}
\item $N_{\Pell}=1$, where the lepton passes the analysis muon or electron selection;
\item maximum track $\BDT>0$;
\item event $\BDT>0$;
\item $\DR(\Pell,\PQjo)>0.4$.
\end{itemize}
Event-level BDT classifiers are used to select signal candidate events in the lepton + exclusive track categories while rejecting background events. The output score of each BDT is used to define both the SRs and CRs.
Separate BDTs are trained for each lepton flavor and for each detector configuration, resulting in a total of four BDTs in the exclusive track category.
These classifiers take as input a similar set of variables as that listed in Table~\ref{tab:dimuon-bdt-variables}, substituting $\Pell_{2}$ properties with equivalent track properties.
The invariant mass of the track-lepton system ranks lowest, although one-dimensional distributions indicate that it still provides substantial discriminating power.
The BDT captures the mass-related information encoded in the other lepton and track features.
Thirteen SRs are defined in the output of the event BDT for each detector configuration and lepton category, with 12 intervals of width 0.05 from 0 to 0.6, and a single SR with $\BDT>0.6$.

\section{Backgrounds}
\label{sec:background-estimation}

The most significant backgrounds consist of events with leptons originating either from jets or from decays of electroweak bosons, primarily arising from \wjets and \zjets processes.
In both cases, at least one of the leptons is nonprompt and originates from jet activity.
These backgrounds are grouped into two main categories, and a dedicated method is developed to estimate the contribution from each.
The measured contributions of the backgrounds in the sideband regions are then used to model the background in the SRs using a maximum likelihood fit.

\subsection{Background in dimuon category}

The dominant background in the dimuon category arises from lepton-track pairs produced in association with jets, primarily from leptons originating in the electroweak decays of hadrons.
This \textit{Jetty} background is estimated using an isolation sideband CR defined by inverting the jet-based isolation criterion on the leptons.
This region is used to extract a template of the BDT score distribution that is consistent with the shape of the Jetty background in the SRs.
Although most leptons are produced within the cores of jets, the distribution of angular distance between the lepton and the jet exhibits a slowly falling tail that extends into the SR.
A second CR, defined by the sideband of the event-level BDT score ($\BDT<0$), is used to normalize the background rate, accounting for differences in the Jetty background production rate between the isolation sideband and main band.
The SR is defined by requiring $\BDT>0$ in the isolation main band.
The predicted Jetty background in the SR is given by:
\ifthenelse{\boolean{cms@external}}{
\begin{equation}
\begin{aligned}
    N^{\text{SR}}_{\text{Jetty}}(x)
        &= \widehat{\text{TF}}_{\text{Jetty}}\, N^{\text{SR}}_{\text{sideband}}(x) \\
        &= \frac{N^{\text{norm CR}}_{\text{main band}}}
                {N^{\text{norm CR}}_{\text{sideband}}}
           N^{\text{SR}}_{\text{sideband}}(x)
\end{aligned}
\end{equation}
}{
\begin{equation}
    N^{\text{SR}}_{\text{Jetty}}(x)
        = \widehat{\text{TF}}_{\text{Jetty}}\, N^{\text{SR}}_{\text{sideband}}(x)
        = \frac{N^{\text{norm CR}}_{\text{main band}}}
                {N^{\text{norm CR}}_{\text{sideband}}}
          N^{\text{SR}}_{\text{sideband}}(x)
\end{equation}
}
where $x$ is the binned BDT output.
The ratio of the yields in the normalization regions is referred to as the transfer factor, $\widehat{\text{TF}}_{\text{Jetty}}$.
The corresponding values are listed in Table~\ref{tab:transfer-factors}.

\begin{figure}[!ht]
\centering
\includegraphics[width=0.45\textwidth]{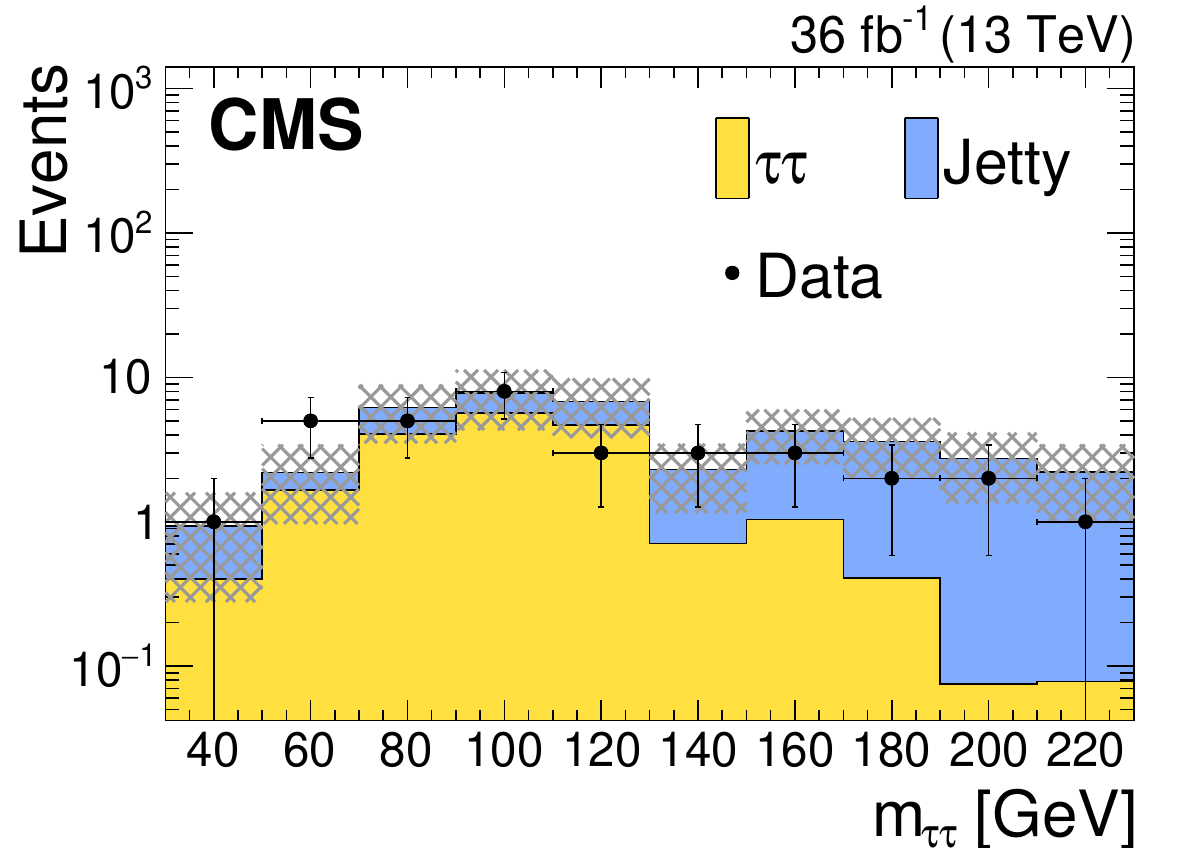}%
\hspace{0.05\textwidth}%
\includegraphics[width=0.45\textwidth]{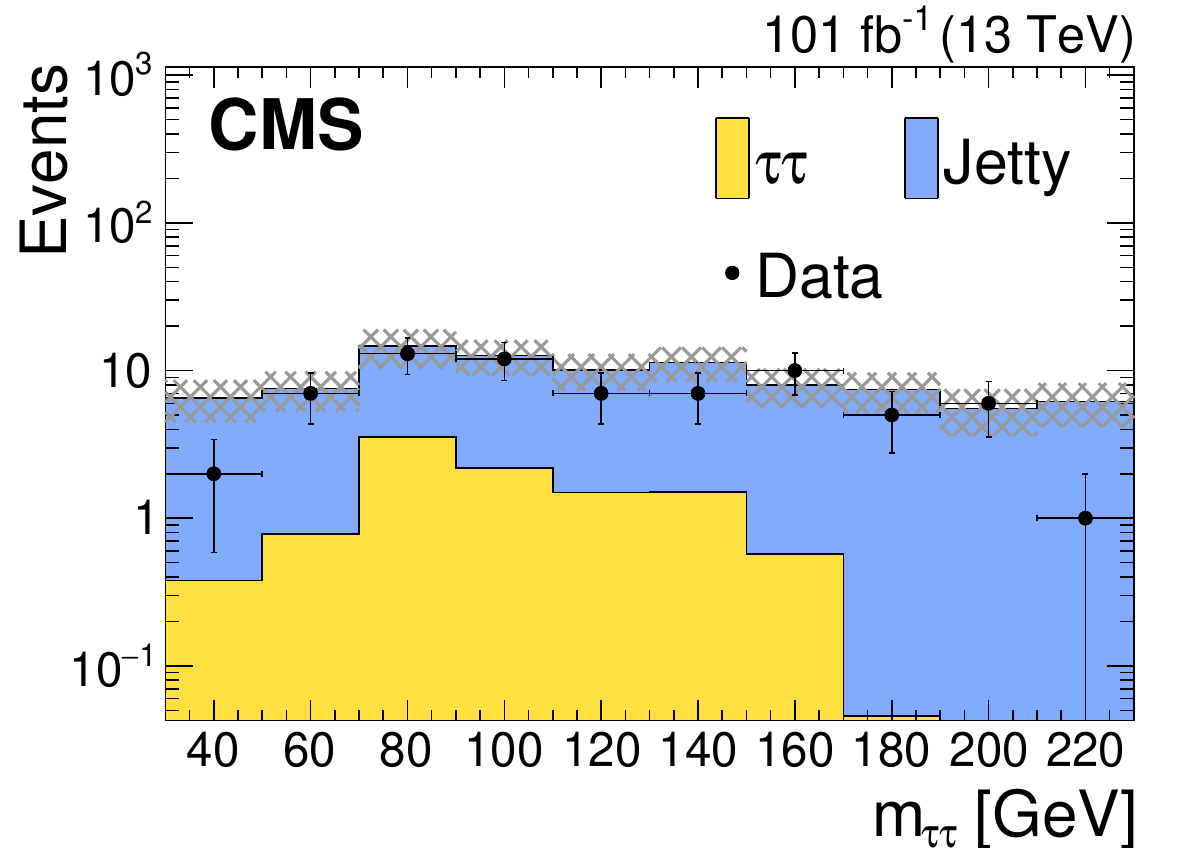}
\caption{Distributions of the reconstructed ditau invariant mass (\mtautau) in the BDT sideband CR, shown for \Phase0 (\cmsLeft) and \Phase1 (\cmsRight) detector configurations. The non-\tautau background is estimated using the data-driven Jetty background method described in the text. The gray hatching shows the statistical uncertainty in the background prediction.}
\label{fig:mtautau-distributions}
\end{figure}

A small contribution to the dimuon category arises from prompt, isolated leptons originating from the \ztautau process.
This background is estimated using simulated event samples corrected with normalization factors derived in a data CR enriched in \PZGammaStar events.
The CR selects events with two signal candidate leptons and requires the reconstructed ditau mass \mtautau to lie within a window of 40--130\GeV, consistent with the \PZ boson mass.
The \mtautau observable is based on the \emph{collinear approximation}, first described in Ref.~\cite{ELLIS1988221_first_mtautau} and employed in previous analyses, \eg, Refs.~\cite{ATLAS:2009zsq,CMS:2007sch}.
In this approximation, it is assumed that each \PGt lepton from the \PZGammaStar decay is sufficiently boosted such that its own decay products are collinear, and that the only source of missing transverse momentum is the neutrinos from the \PGt lepton decays.
The visible muon momenta, together with \ptmiss and the known \PGt lepton mass, are used to reconstruct the \PGt four-vectors and calculate \mtautau.
The data and background prediction are shown in the \tautau CR in Fig.~\ref{fig:mtautau-distributions}.
The ratio of data to simulation in this region is used to extract the correction factors, denoted as $\widehat{\text{TF}}_{\tautau}$, which are reported in Table~\ref{tab:transfer-factors}.
Contamination from the Jetty background is subtracted from data using the method described above. The uncertainty in the normalization correction is large in \Phase1 because of the large rate of Jetty background in the TF measurement region.

\begin{table}[!ht]
\centering
\topcaption{Transfer factors and their associated statistical uncertainties used to extrapolate background predictions from CRs to the SR.}
\label{tab:transfer-factors}
\renewcommand{\arraystretch}{1.1}
\begin{scotch}{lccccc}
Method & Flavor & Phase & $\widehat{\text{TF}}$ & Stat.\ unc. & Rel.\ unc. \\ \hline
Jetty & Muon & 0 & 0.73 & 0.14 & 19\% \\
Jetty & Muon & 1 & 0.62 & 0.06 & 9\% \\
Exclusive track & Muon & 0 & 1.11 & 0.04 & 4\% \\
Exclusive track & Muon & 1 & 1.07 & 0.02 & 2\% \\
Exclusive track & Electron & 0 & 1.04 & 0.05 & 5\% \\
Exclusive track & Electron & 1 & 1.05 & 0.03 & 3\% \\
\tautau & Muon & 0 & 1.18 & 0.45 & 38\% \\
\tautau & Muon & 1 & 0.29 & 0.26 & 90\% \\
\end{scotch}
\end{table}

\subsection{Background in exclusive track category}

While tracks in signal candidate events are required to have the opposite charge from the identified lepton, background events featuring same-charge tracks are otherwise kinematically similar in the overwhelming majority of cases.
Backgrounds involving isolated leptons from the prompt interaction are found to contribute negligibly to the SRs.
Therefore, the background in the SRs of the exclusive track category is estimated using the same-charge CR, defined by inverting the opposite-charge requirement while applying the full analysis selection.
The distribution in the same-charge data CR is adjusted with a small normalization correction derived using events with $\BDT<0$ to match the count in the opposite-sign region.
The method is validated through closure tests performed in simulation, comparing the prediction from the above method to the expected result obtained directly from simulation.
The resulting transfer factors are reported in Table~\ref{tab:transfer-factors}.

\subsection{Systematic uncertainty}

Systematic uncertainty in the background prediction is assessed based on discrepancies between observed and predicted event counts in a same-charge dilepton CR, as well as from validation studies performed using simulation.
The Jetty background estimation method is performed in simulated events and compared with the results from direct simulation, and a first-order polynomial is fit to the ratio in bins of the classifier.
The best-fit result is found to be statistically consistent with the unit line.
The fit parameters are varied by their statistical uncertainty to yield an alternate prediction, the difference of which from the nominal prediction is propagated as a systematic uncertainty.
Uncertainty values range from 8 to 22\% of the estimated yields, increasing with the value of the event classifier, and are treated as fully correlated across bins.
An uncertainty of 100\% is assigned to the \ztautau background estimates to cover the large variation in scale factors resulting from the subtraction of the Jetty background from the \ztautau CR; this assignment has negligible impact on the sensitivity.
Several sources of uncertainty affecting the signal yield are also identified and estimated, including uncertainty in the modeling of the jet energy response, the \pt spectrum of ISR, the efficiency of lepton and \PQb-tagged jet reconstruction, identification, and selection, as well as the integrated luminosity, trigger efficiency, and pileup profile.
These uncertainties amount to a total of 5--20\% in the SRs and are incorporated into a maximum likelihood fit.
Each source of systematic uncertainty is log-normal distributed, with a corresponding nuisance parameter that modifies the predicted rate of a given process via a multiplicative factor.
The width of each log-normal distribution reflects the relative variation in the predicted yield under a one-standard-deviation ($\sigma$) shift of the associated uncertainty.
Statistical uncertainties in CRs and simulated event yields are incorporated using gamma-distributed nuisance parameters.
The likelihood combines all SRs in all event categories simultaneously.

\section{Results and interpretation}
\label{sec:results}

\begin{figure*}[!pt]
\centering
\includegraphics[width=0.45\textwidth]{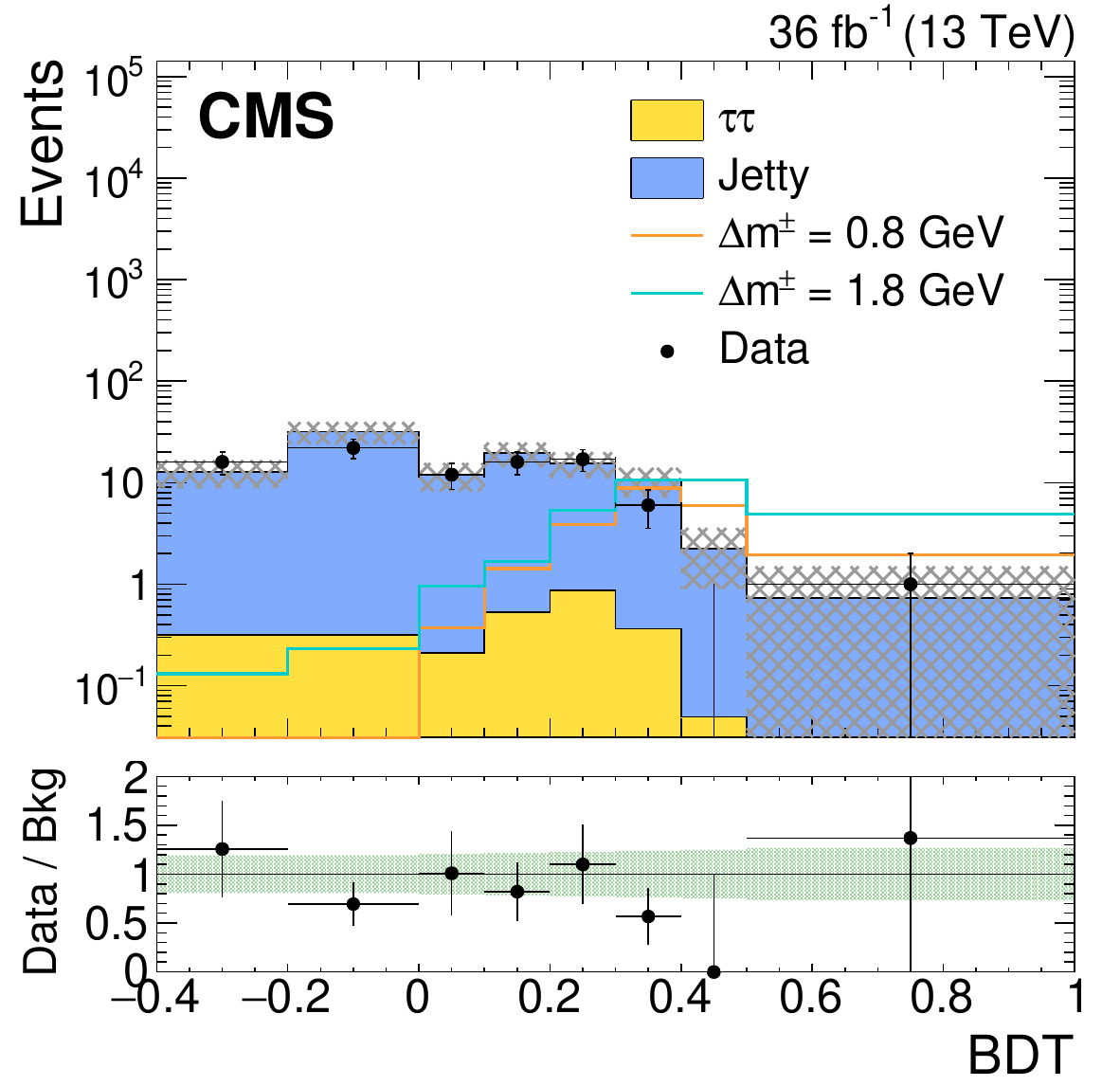}%
\hspace*{0.05\textwidth}%
\includegraphics[width=0.45\textwidth]{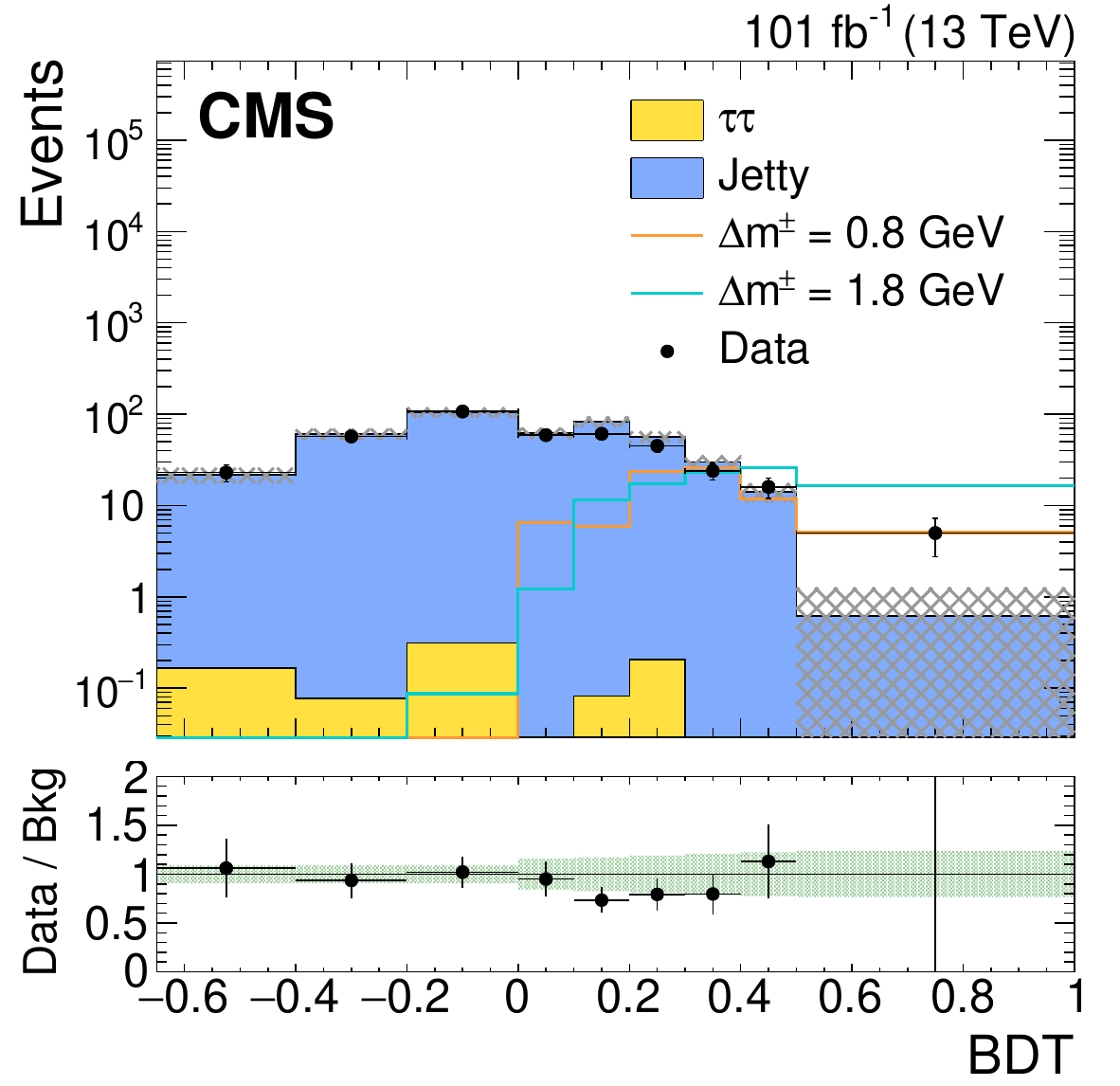} \\
\includegraphics[width=0.45\textwidth]{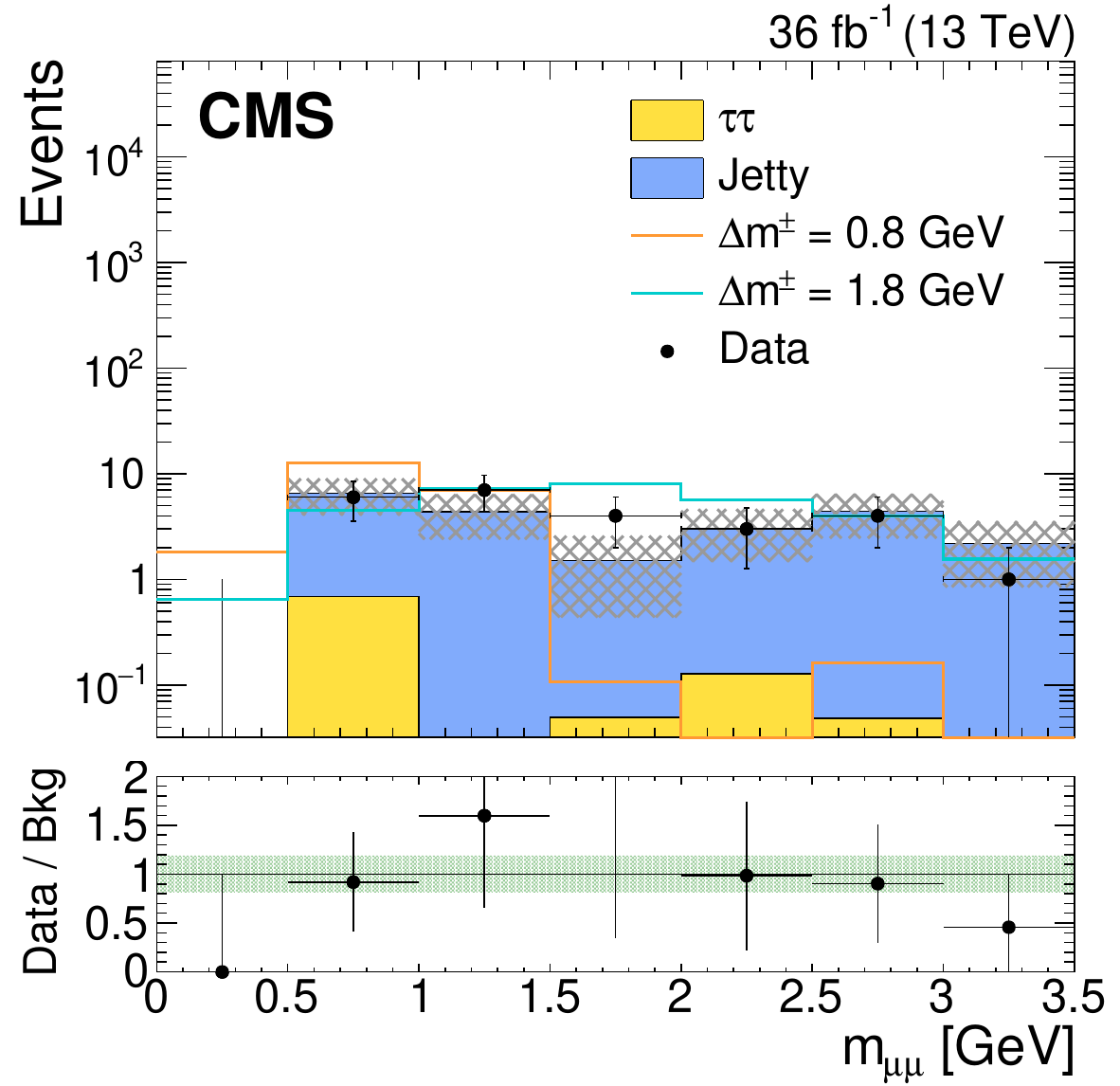}%
\hspace*{0.05\textwidth}%
\includegraphics[width=0.45\textwidth]{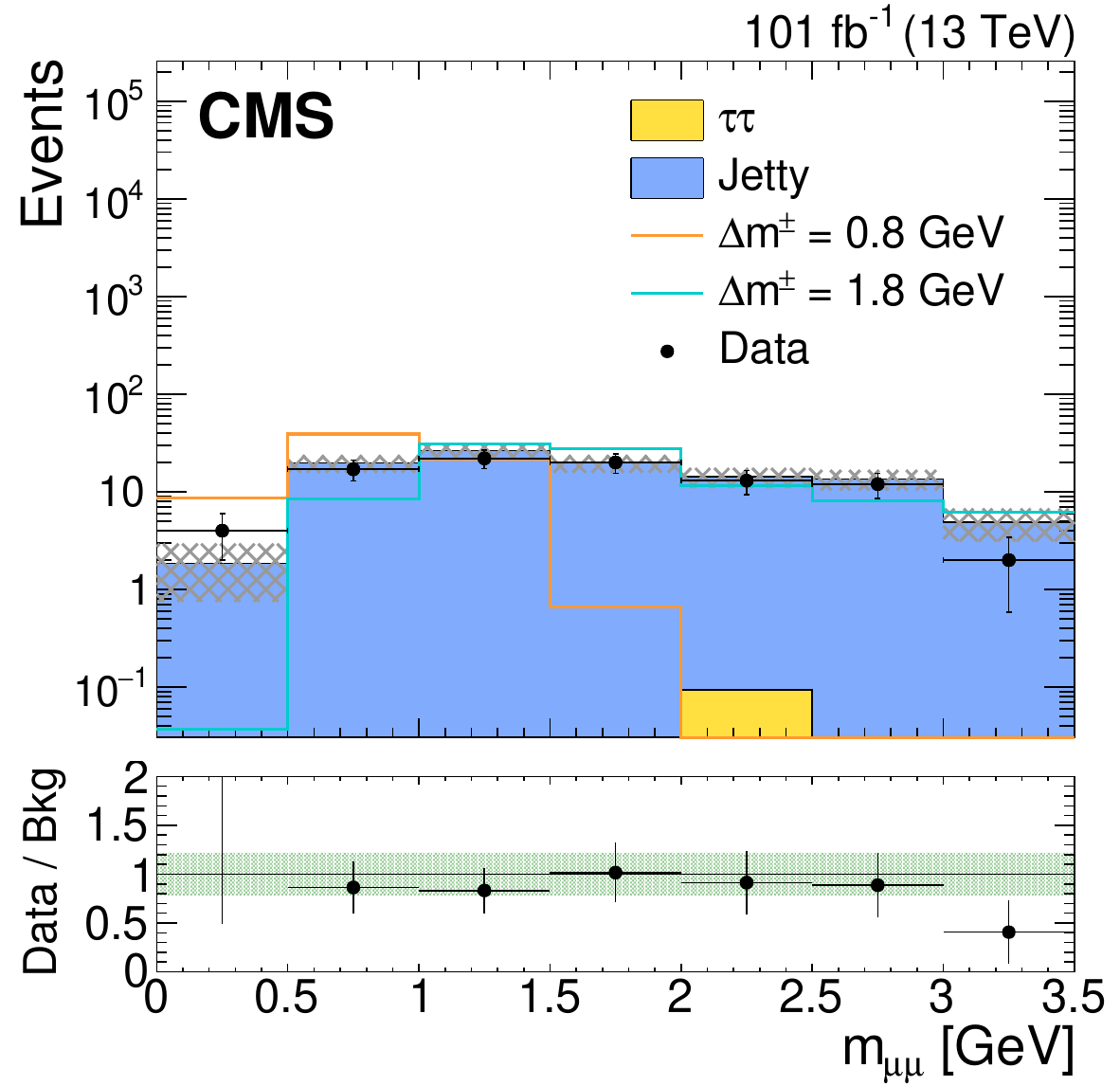}
\caption{Prefit expected and observed distributions of the event BDT output score in the SRs for the dimuon category (upper) and of the dimuon invariant mass for events with event classifier scores greater than 0.1 (lower), shown separately for \Phase0 (left) and \Phase1 (right) configurations.
The gray hatching shows the statistical uncertainty in the background prediction. In the lower panel, the green band indicates the relative systematic uncertainty in the predicted background, while the black bars represent the total uncertainty, including both statistical and systematic components.
Two example signal scenarios are also shown as colored lines.}
\label{fig:data-driven-full-background-prediction}
\end{figure*}

\begin{figure*}[!pt]
\centering
\includegraphics[width=0.45\textwidth]{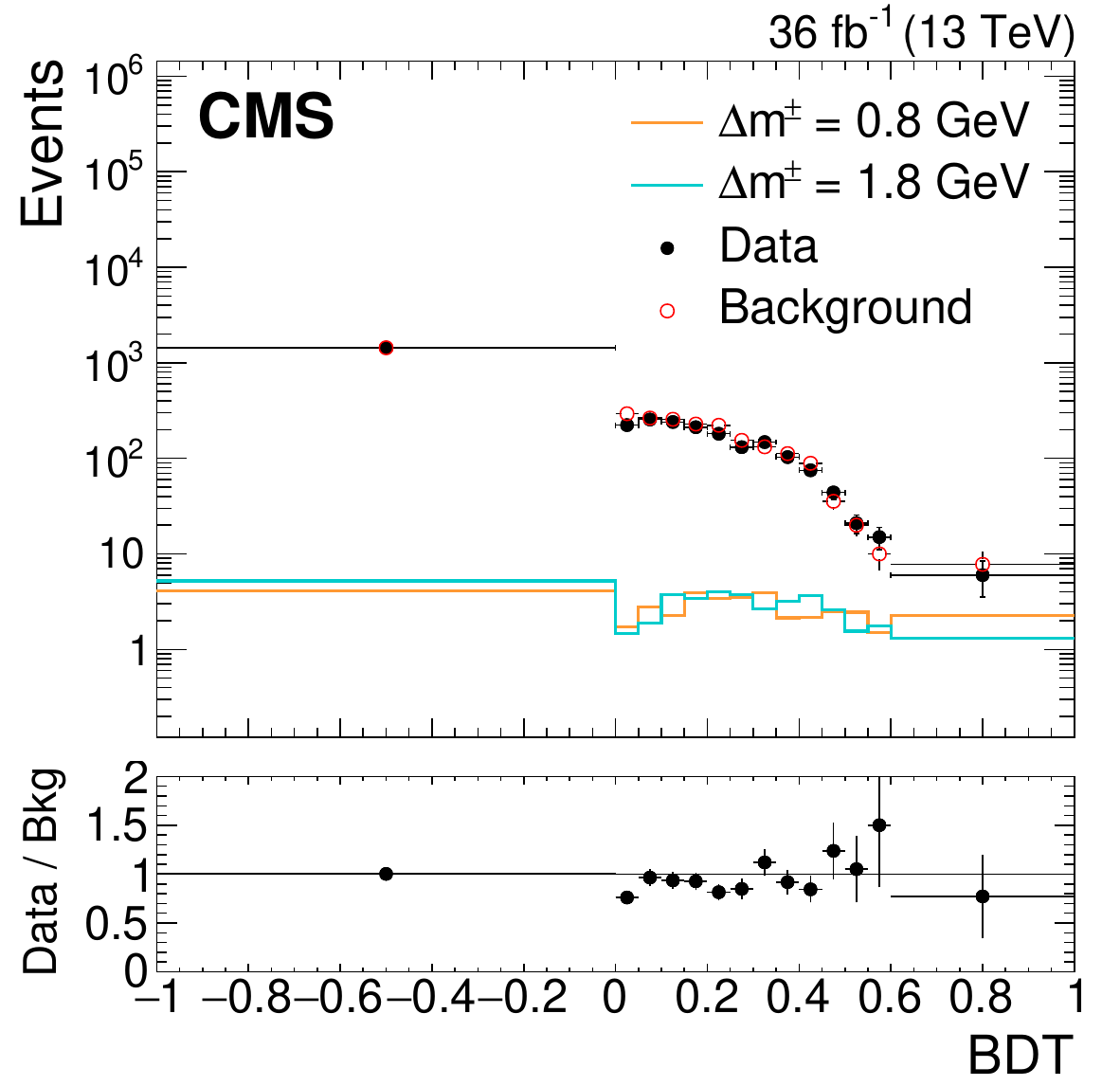}%
\hspace*{0.05\textwidth}%
\includegraphics[width=0.45\textwidth]{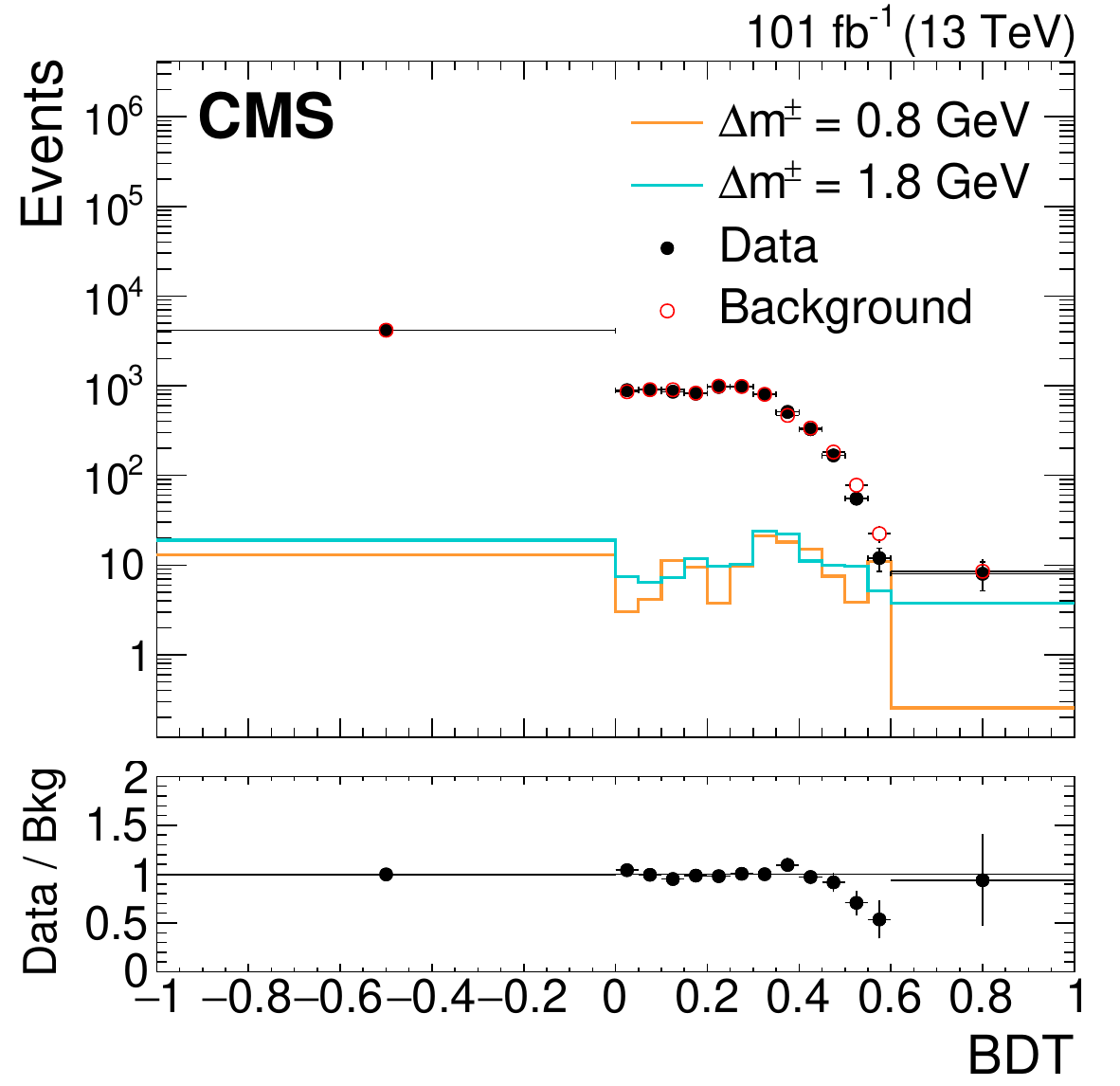} \\
\includegraphics[width=0.45\textwidth]{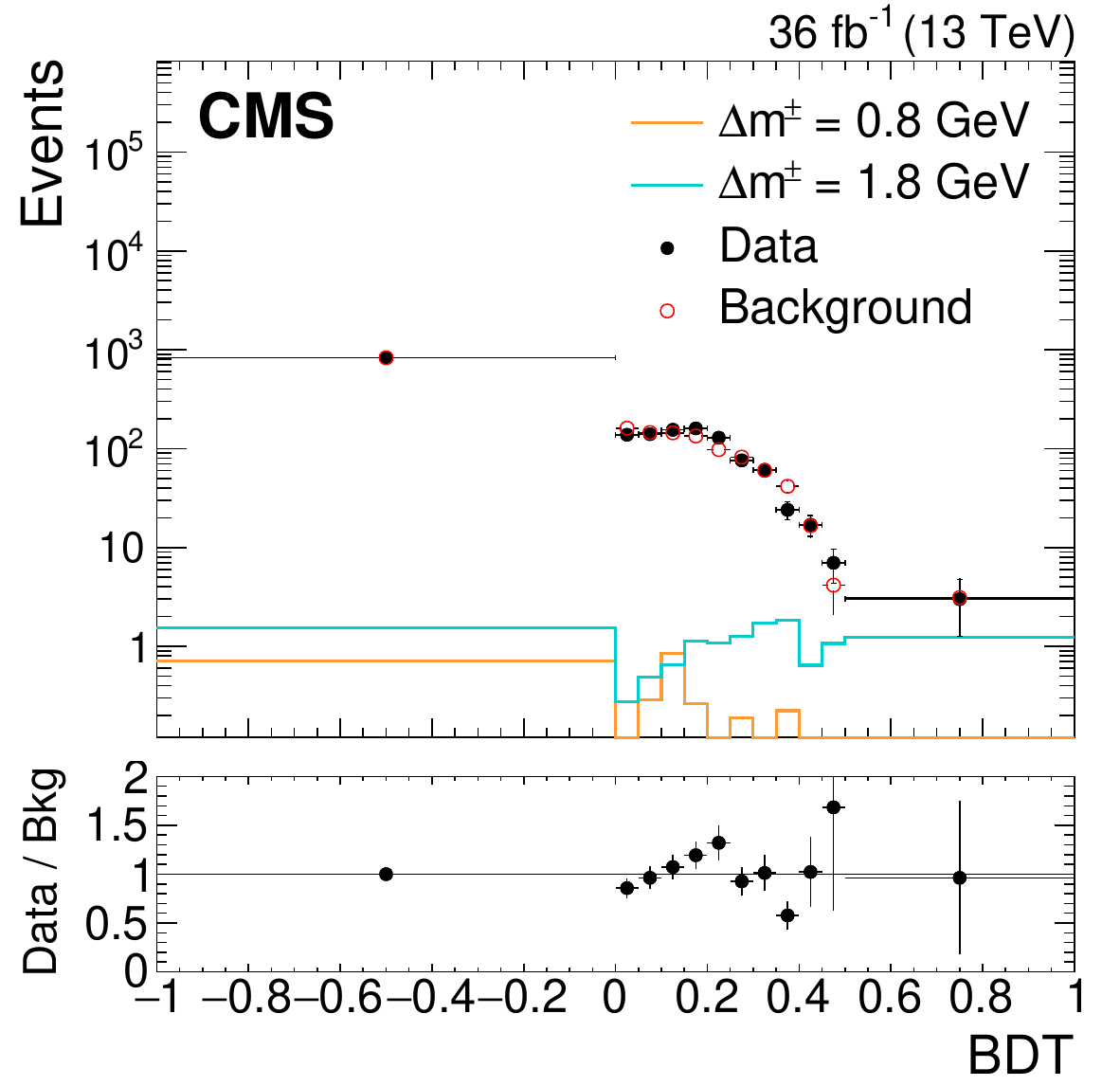}%
\hspace*{0.05\textwidth}%
\includegraphics[width=0.45\textwidth]{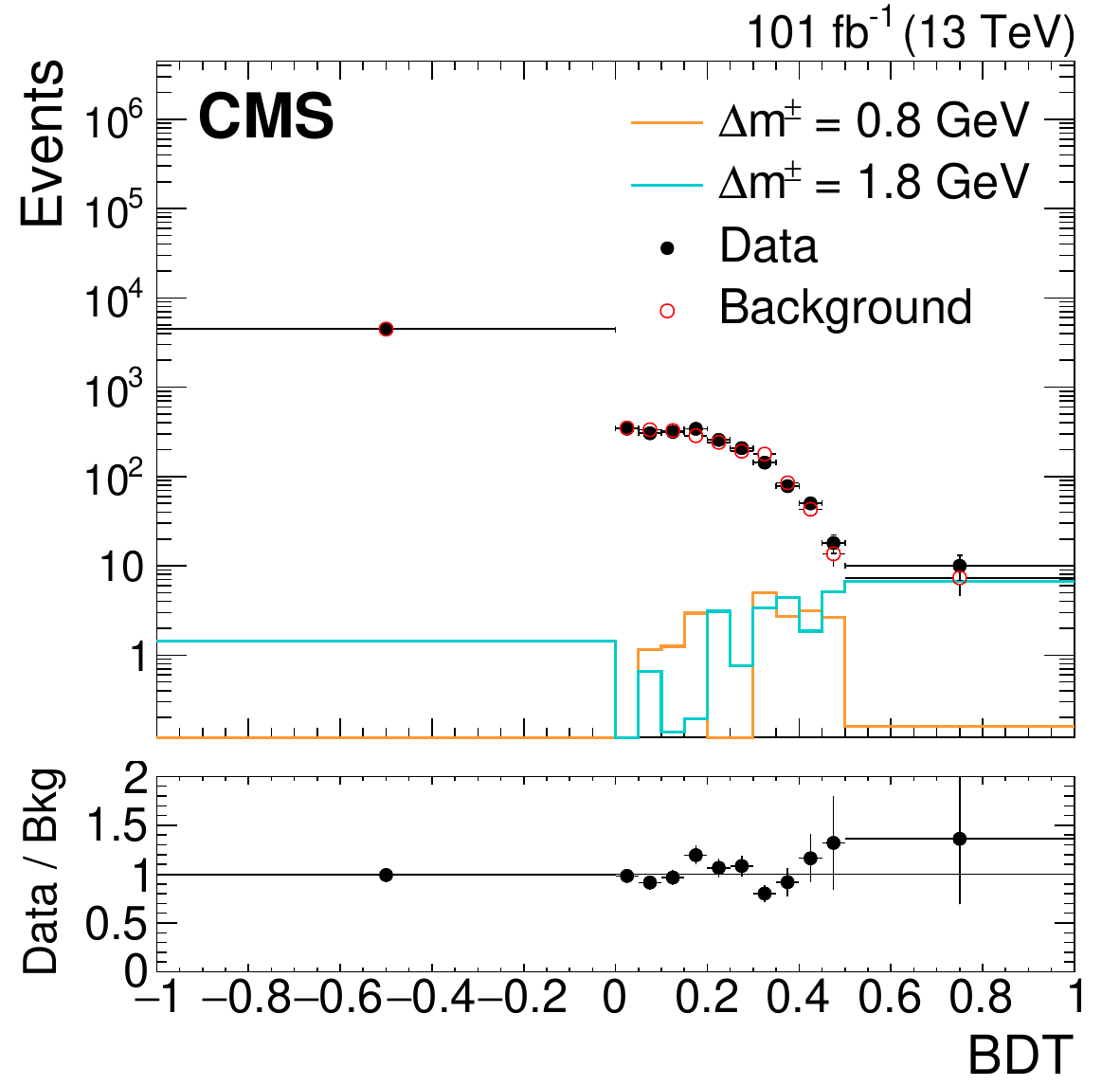}
\caption{Prefit expected and observed distributions of the event BDT output score in the SRs for the muon + exclusive track category (upper), and the electron + exclusive track category (lower), shown separately for \Phase0 (left) and \Phase1 (right).
In the lower panel, the vertical black bars represent the total uncertainty, including both statistical and systematic components.
Example signal benchmark scenarios are also shown as colored lines.}
\label{fig:data-driven-full-background-prediction-exc}
\end{figure*}

The results of the analysis are shown in Fig.~\ref{fig:data-driven-full-background-prediction} for the dimuon category and in Fig.~\ref{fig:data-driven-full-background-prediction-exc} for the lepton + exclusive track categories.
No significant deviation from the SM expectation is observed, and the background model provides a good description of the data.
A small excess is observed in the most sensitive SR of the dimuon category, more pronounced in the \Phase1 data set, as seen in Fig.~\ref{fig:data-driven-full-background-prediction}.
The corresponding local significance, accounting for all SRs from both data sets, is 2.2 standard deviations.

The results are interpreted in the context of the compressed Higgsino simplified model introduced in Section~\ref{sec:search-introduction}, using the maximum likelihood fit.
Both the expected and observed limits are derived using the asymptotic approximation in a maximum likelihood framework~\cite{Cowan:2010js}.
Observed and expected counts are incorporated into the likelihood for all SRs, corresponding to bins in the results histograms with $\BDT>0$.
The \CLs method~\cite{Junk:1999kv,A_L_Read_2002} is employed to compute exclusion limits at the 95\% confidence level (\CL)~\cite{CMS:2024onh}.

\begin{figure}[t!h]
\centering
\includegraphics[width=\cmsFigWidth]{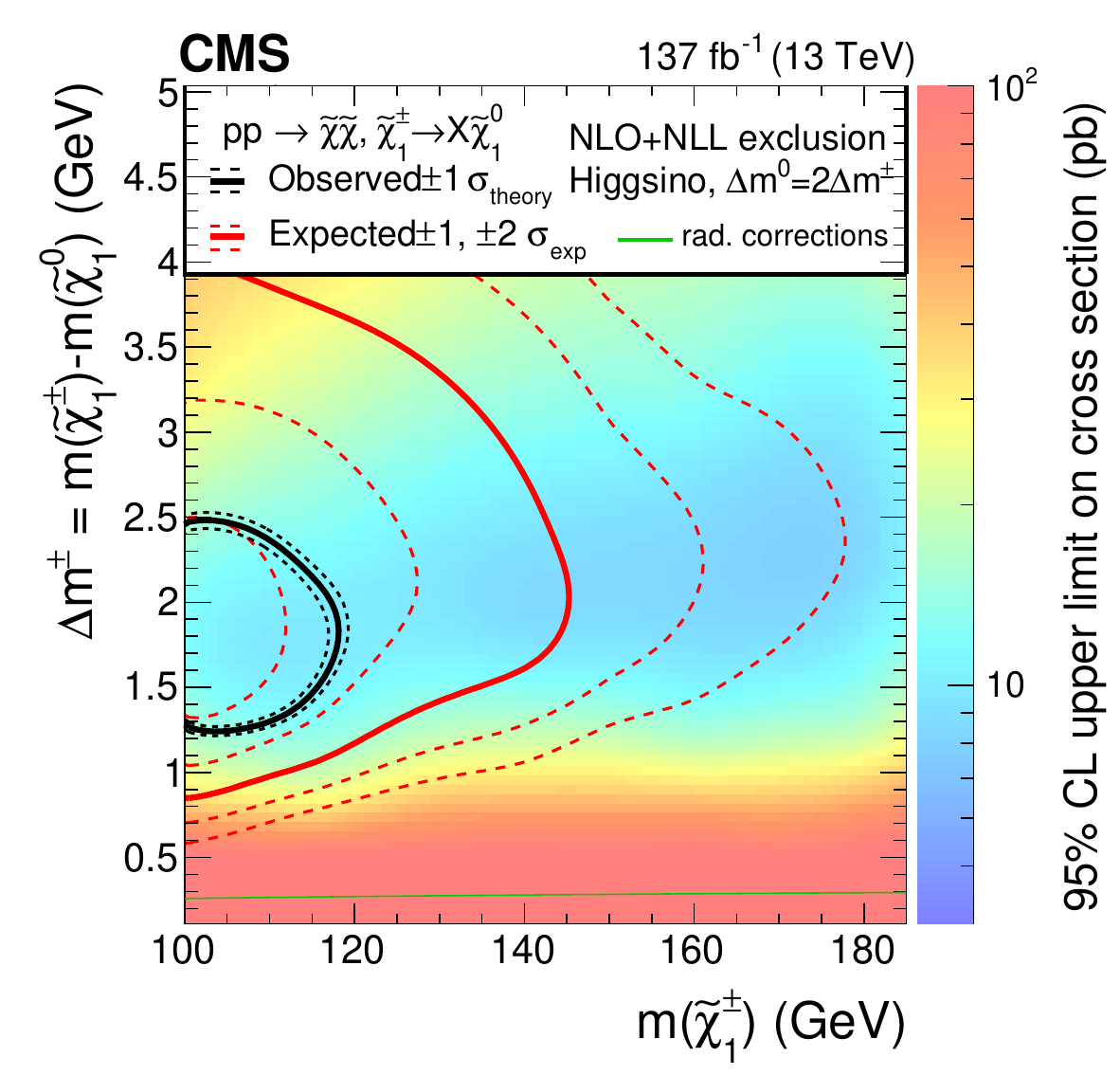}
\caption{The 95\% CL upper limits on the fully degenerate Higgsino production cross section, calculated at NLO+NLL precision~\cite{Fuks:2012qx,Fuks:2013vua}, based on all analysis channels studied in this paper.
All relevant production modes are simulated at LO, and the \PZstar boson is set to decay into either two electrons or two muons with a branching fraction of 5\%.
The expected (red) and observed (black) exclusion contours are shown assuming the theoretical cross section.
Dashed red lines indicate the expected limits with $\pm$1 and $\pm$2 standard deviation experimental uncertainties.
Dashed black lines indicate the observed limit when varying the theoretical cross section by its uncertainty.
The green line represents the minimum \Dmpm allowed by the theoretical calculation accounting for radiative corrections, as described in Ref.~\cite{Nagata:2014wma}.}
\label{fig:limit}
\end{figure}

The limits are presented in the plane of \Dmpm and $m(\PSGcpmDo)$ in Fig.~\ref{fig:limit}.
As discussed in Section~\ref{sec:search-introduction}, the mass splitting between the neutralinos satisfies $\Dmz=2\Dmpm$, consistent with scenarios of large $\tan\beta$.
The region enclosed by the observed contour is excluded at 95\% CL, while the color scale indicates the corresponding upper limits on the signal cross section.
The green curve represents the minimum allowed value of \Dmpm from theoretical calculations that include radiative corrections, following the treatment in Ref.~\cite{Nagata:2014wma}.
The sensitivity peaks around $\Dmpm\approx2\GeV$, where the analysis probes charginos up to masses of about 145\GeV, with the observed exclusion reaching up to approximately 115\GeV.
The observed limit is weaker than the expected limit due to the somewhat larger observed count in the tightest SR of \Phase1.

\section{Summary}
\label{sec:summary}

A search for Higgsino pair production in compressed mass spectra scenarios is performed using low-momentum dimuon and lepton-track pairs in proton-proton collisions at $\sqrt{s}=13\TeV$, based on a data sample corresponding to an integrated luminosity of 137\fbinv~\cite{CMS:LUM-17-003,CMS-PAS-LUM-17-004,CMS-PAS-LUM-18-002} collected in 2016, 2017, and 2018 with the CMS detector.
The results are interpreted in a simplified model featuring a dark matter candidate neutralino \PSGczDo that is nearly mass-degenerate with a slightly heavier neutralino \PSGczDt and charginos \PSGcpmDo.
The search targets a region of parameter space where sensitivity was limited in previous analyses~\cite{Aad_2020,CMS:2021edw}, and extends the reach in \Dmz ($=2\Dmpm$) by 0.5--2\GeV in the most compressed phase space.
This region, characterized by low-mass Higgsino eigenstates, is of particular theoretical interest because of its relevance for naturalness and fine-tuning arguments~\cite{Martin:1997ns, BARBIERI198863, de_Carlos_1993}, addressing both the large and small hierarchy problems. Values of \Dmz are probed down to 1.5\GeV, assuming a chargino mass of 100\GeV. The maximum excluded Higgsino mass is 115\GeV, corresponding to a \Dmz of 3.5\GeV. These results place stringent constraints on natural supersymmetry and other models predicting electroweak multiplet dark matter.

\begin{acknowledgments}
We congratulate our colleagues in the CERN accelerator departments for the excellent performance of the LHC and thank the technical and administrative staffs at CERN and at other CMS institutes for their contributions to the success of the CMS effort. In addition, we gratefully acknowledge the computing centers and personnel of the Worldwide LHC Computing Grid and other centers for delivering so effectively the computing infrastructure essential to our analyses. Finally, we acknowledge the enduring support for the construction and operation of the LHC, the CMS detector, and the supporting computing infrastructure provided by the following funding agencies: SC (Armenia), BMBWF and FWF (Austria); FNRS and FWO (Belgium); CNPq, CAPES, FAPERJ, FAPERGS, and FAPESP (Brazil); MES and BNSF (Bulgaria); CERN; CAS, MoST, and NSFC (China); MINCIENCIAS (Colombia); MSES and CSF (Croatia); RIF (Cyprus); SENESCYT (Ecuador); ERC PRG, TARISTU24-TK10 and MoER TK202 (Estonia); Academy of Finland, MEC, and HIP (Finland); CEA and CNRS/IN2P3 (France); SRNSF (Georgia); BMFTR, DFG, and HGF (Germany); GSRI (Greece); NKFIH (Hungary); DAE and DST (India); IPM (Iran); SFI (Ireland); INFN (Italy); MSIT and NRF (Republic of Korea); MES (Latvia); LMTLT (Lithuania); MOE and UM (Malaysia); BUAP, CINVESTAV, CONACYT, LNS, SEP, and UASLP-FAI (Mexico); MOS (Montenegro); MBIE (New Zealand); PAEC (Pakistan); MES, NSC, and NAWA (Poland); FCT (Portugal); MESTD (Serbia); MICIU/AEI and PCTI (Spain); MOSTR (Sri Lanka); Swiss Funding Agencies (Switzerland); MST (Taipei); MHESI (Thailand); TUBITAK and TENMAK (T\"{u}rkiye); NASU (Ukraine); STFC (United Kingdom); DOE and NSF (USA).

\hyphenation{Rachada-pisek} Individuals have received support from the Marie-Curie program and the European Research Council and Horizon 2020 Grant, contract Nos.\ 675440, 724704, 752730, 758316, 765710, 824093, 101115353, 101002207, 101001205, and COST Action CA16108 (European Union); the Leventis Foundation; the Alfred P.\ Sloan Foundation; the Alexander von Humboldt Foundation; the Science Committee, project no. 22rl-037 (Armenia); the Fonds pour la Formation \`a la Recherche dans l'Industrie et dans l'Agriculture (FRIA) and Fonds voor Wetenschappelijk Onderzoek contract No. 1228724N (Belgium); the Beijing Municipal Science \& Technology Commission, No. Z191100007219010, the Fundamental Research Funds for the Central Universities, the Ministry of Science and Technology of China under Grant No. 2023YFA1605804, the Natural Science Foundation of China under Grant No. 12061141002, 12535004, and USTC Research Funds of the Double First-Class Initiative No.\ YD2030002017 (China); the Ministry of Education, Youth and Sports (MEYS) of the Czech Republic; the Shota Rustaveli National Science Foundation, grant FR-22-985 (Georgia); the Deutsche Forschungsgemeinschaft (DFG), among others, under Germany's Excellence Strategy -- EXC 2121 ``Quantum Universe" -- 390833306, and under project number 400140256 - GRK2497; the Hellenic Foundation for Research and Innovation (HFRI), Project Number 2288 (Greece); the Hungarian Academy of Sciences, the New National Excellence Program - \'UNKP, the NKFIH research grants K 131991, K 133046, K 138136, K 143460, K 143477, K 146913, K 146914, K 147048, 2020-2.2.1-ED-2021-00181, TKP2021-NKTA-64, and 2021-4.1.2-NEMZ\_KI-2024-00036 (Hungary); the Council of Science and Industrial Research, India; ICSC -- National Research Center for High Performance Computing, Big Data and Quantum Computing, FAIR -- Future Artificial Intelligence Research, and CUP I53D23001070006 (Mission 4 Component 1), funded by the NextGenerationEU program (Italy); the Latvian Council of Science; the Ministry of Education and Science, project no. 2022/WK/14, and the National Science Center, contracts Opus 2021/41/B/ST2/01369, 2021/43/B/ST2/01552, 2023/49/B/ST2/03273, and the NAWA contract BPN/PPO/2021/1/00011 (Poland); the Funda\c{c}\~ao para a Ci\^encia e a Tecnologia, grant CEECIND/01334/2018 (Portugal); the National Priorities Research Program by Qatar National Research Fund;  MICIU/AEI/10.13039/501100011033, ERDF/EU, "European Union NextGenerationEU/PRTR", and Programa Severo Ochoa del Principado de Asturias (Spain); the Chulalongkorn Academic into Its 2nd Century Project Advancement Project, the National Science, Research and Innovation Fund program IND\_FF\_68\_369\_2300\_097, and the Program Management Unit for Human Resources \& Institutional Development, Research and Innovation, grant B39G680009 (Thailand); the Kavli Foundation; the Nvidia Corporation; the SuperMicro Corporation; the Welch Foundation, contract C-1845; and the Weston Havens Foundation (USA).
\end{acknowledgments}\section*{Data availability} Release and preservation of data used by the CMS Collaboration as the basis for publications is guided by the  \href{https://doi.org/10.7483/OPENDATA.CMS.1BNU.8V1W}{CMS data preservation, re-use and open access policy}.

\bibliography{auto_generated}
\cleardoublepage \appendix\section{The CMS Collaboration \label{app:collab}}\begin{sloppypar}\hyphenpenalty=5000\widowpenalty=500\clubpenalty=5000\cmsinstitute{Yerevan Physics Institute, Yerevan, Armenia}
{\tolerance=6000
A.~Hayrapetyan, V.~Makarenko\cmsorcid{0000-0002-8406-8605}, A.~Tumasyan\cmsAuthorMark{1}\cmsorcid{0009-0000-0684-6742}
\par}
\cmsinstitute{Institut f\"{u}r Hochenergiephysik, Vienna, Austria}
{\tolerance=6000
W.~Adam\cmsorcid{0000-0001-9099-4341}, J.W.~Andrejkovic, L.~Benato\cmsorcid{0000-0001-5135-7489}, T.~Bergauer\cmsorcid{0000-0002-5786-0293}, M.~Dragicevic\cmsorcid{0000-0003-1967-6783}, C.~Giordano, P.S.~Hussain\cmsorcid{0000-0002-4825-5278}, M.~Jeitler\cmsAuthorMark{2}\cmsorcid{0000-0002-5141-9560}, N.~Krammer\cmsorcid{0000-0002-0548-0985}, A.~Li\cmsorcid{0000-0002-4547-116X}, D.~Liko\cmsorcid{0000-0002-3380-473X}, M.~Matthewman, I.~Mikulec\cmsorcid{0000-0003-0385-2746}, J.~Schieck\cmsAuthorMark{2}\cmsorcid{0000-0002-1058-8093}, R.~Sch\"{o}fbeck\cmsAuthorMark{2}\cmsorcid{0000-0002-2332-8784}, D.~Schwarz\cmsorcid{0000-0002-3821-7331}, M.~Shooshtari\cmsorcid{0009-0004-8882-4887}, M.~Sonawane\cmsorcid{0000-0003-0510-7010}, W.~Waltenberger\cmsorcid{0000-0002-6215-7228}, C.-E.~Wulz\cmsAuthorMark{2}\cmsorcid{0000-0001-9226-5812}
% ...
\par}
\vskip\cmsinstskip
\dag:~Deceased\\
$^{1}$Also at Yerevan State University, Yerevan, Armenia\\
$^{2}$Also at TU Wien, Vienna, Austria\\
% ...
\end{sloppypar}
%%% END EDITABLE REGION %%%
% skeleton_end
\end{document}